\documentclass[doublecol, figures]{epl2} 
\usepackage{blindtext}
\usepackage{subcaption}
\usepackage{amsmath}
\usepackage{color} 
\title{Partial synchronization in empirical brain networks as a model for unihemispheric sleep}
\shorttitle{Title} %Insert here a short version of the title if it exceeds 70 characters

\author{Lukas Ramlow\inst{1} \and Jakub Sawicki\inst{1} \and Anna Zakharova\inst{1} \and Jaroslav Hlinka\inst{2,3} \and Jens Christian Claussen\inst{4} \and Eckehard Sch{\"o}ll \inst{1}}
\shortauthor{L. Ramlow \etal}

\institute{                    
  \inst{1} Institut f{\"u}r Theoretische Physik, Technische Universit{\"a}t Berlin, Hardenbergstra\ss{}e 36, 10623 Berlin, Germany \\
  \inst{2} Institute of Computer Science of the Czech Academy of Sciences, Pod Vodarenskou vezi 2, 18207 Prague 8, Czech Republic \\
  \inst{3} National Institute of Mental Health, Topolov\'{a} 748, 250 67 Klecany, Czech Republic \\
  \inst{4} Department of Mathematics, Aston University, Aston Trinagle, Birmingham B4 ET, United Kingdom
}

\abstract{We analyze partial synchronization patterns in a network of FitzHugh-Nagumo oscillators with empirical structural connectivity measured in healthy human subjects. We report a dynamical asymmetry between the hemispheres, induced by the natural structural asymmetry. We show that the dynamical asymmetry can be enhanced by introducing the inter-hemispheric coupling strength as a control parameter for partial synchronization patterns. We specify the possible modalities for existence of unihemispheric sleep in human brain, where one hemisphere sleeps while the other remains awake. In fact, this state is common among migratory birds and mammals like aquatic species.}

\begin{document}

\maketitle
%-----------------------------INTRODUCTION------------------------------------------------------------------
\section{Introduction}
A well-known phenomenon in nature is unihemispheric slow-wave sleep, exhibited by aquatic mammals including whales, dolphins and seals, and multiple bird species. Unihemispheric sleep, as the name suggests, is the remarkable ability to engage in deep (slow-wave) sleep with a single hemisphere of the brain while the other hemisphere remains awake \cite{RAT00,RAT16,MAS16}. Interestingly, sleep and wakefulness are characterized by a high and low degree of synchronization, respectively \cite{SCH08o}. In the human brain the first-night effect, which describes troubled sleep in a novel environment, has been related to asymmetric dynamics recently, i.e., a manifestation of one hemisphere of the brain being more vigilant than the other \cite{TAM16}. 
Sleep is a dynamical macrostate of the brain that is observed over a wide range of animal species. Sleep is accompanied by a loss of consciousness and conscious perceptions, and muscle activity is reduced or absent. Sleep alternates between rapid-eye-movement (REM) and non-REM stages N1, N2, N3, where the latter are dominated by slow oscillations (1 Hz and below) which can also emerge locally \cite{VOH11,LES11}. Sleep stage switching dynamics includes wake/sleep asymmetric stochasticity \cite{SAL17a}, but obeys an underlying control by regulatory circuits forming bistable biological flipflop switches \cite{SCS01,FUL06,SFP10,BBM12}, and sleep regulation is coupled to the sleep oscillations of the thalamocortical system \cite{SCO16}. While most animals follow a similar qualitative sleep pattern and fall into sleep with both hemispheres, in certain bird and mammal species sleep can be unihemispheric \cite{MAS16}. It has been speculated that unihemispheric sleep is related to the spontaneous symmetry-breaking phenomenon of chimera states in oscillator networks \cite{ABR08,MOT10}; those states combine spatially coexisting domains of synchronized and desynchronized dynamics \cite{KUR02a,ABR04,SHI04,PAN15,SCH16b}.\\
While the neurophysiological processes that ensure the existence of this dynamical state of unihemispheric sleep remain largely unknown, it is presumed that a certain degree of interhemispheric separation is a necessary condition for this pattern to persist. Therefore we propose to model unihemispheric sleep by a two-community network of the two hemispheres where the inter-hemispheric coupling strength is smaller than the intra-hemispheric coupling. We model the spiking dynamics of the neurons by the paradigmatic FitzHugh-Nagumo model, and investigate possible partial synchronization patterns.

%-----------------------------MODEL-------------------------------------------------------------------------
\section{Model}
\begin{figure*}[tp]
    \begin{center}
    \onefigure[width=1.0\linewidth]{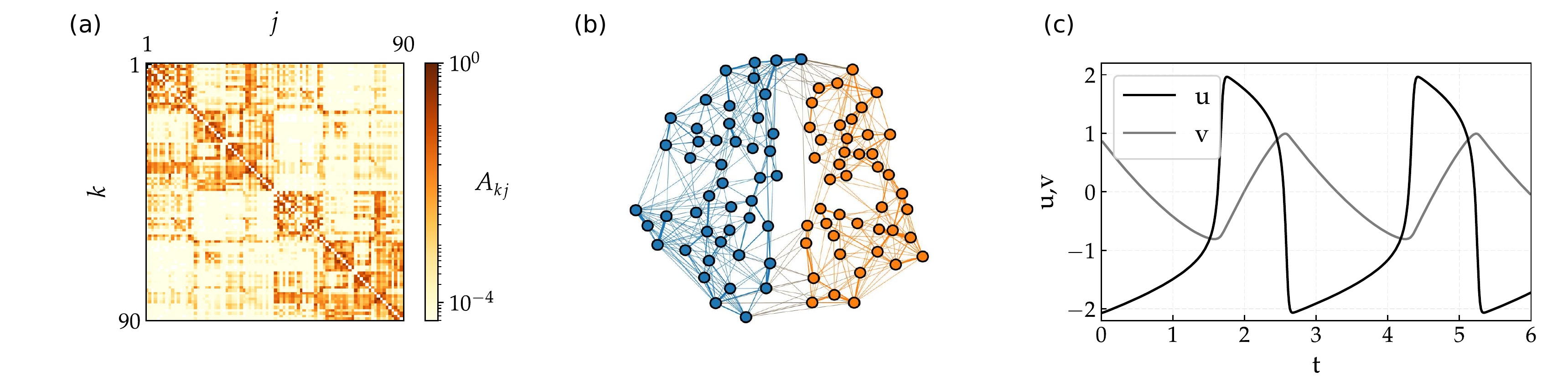}
    \caption{(color online) Model for the hemispheric brain structure: (a) Weighted adjacency matrix $A_{kj}$ of the averaged empirical structural brain network derived from twenty healthy human subjects by averaging over the coupling between two brain regions $k$ and $j$. The brain regions $k,j$ are taken from the Automated Anatomic Labeling atlas \cite{TZO02}, but re-labeled such that $k=1,...,45$ and $k=46,...,90$ correspond to the left and right hemisphere, respectively. (b) Schematic representation of the graph of the brain structure with highlighted left (dark blue) and right (light orange) hemisphere. (c) Dynamics of activator $u$ and inhibitor variable $v$ of the FitzHugh-Nagumo model in the oscillatory regime with $\epsilon = 0.05$ and $a = 0.5$ vs time $t$.}
    \label{fig.1}
    \end{center}
\end{figure*}

%What its all about:
We consider an empirical structural brain network shown in Fig.\,\ref{fig.1} where every region of interest is modeled by a single FitzHugh-Nagumo (FHN) oscillator. \\
%Network:
The brain network was obtained from diffusion-weighted magnetic resonance imaging data measured in healthy human subjects as part of a larger study focusing on connectivity changes in schizophrenia. For details of the measurement procedure including acquisition parameters, see \cite{MEL15}, for previous utilization of the structural networks to analyze chimera states see \cite{CHO18}. The data were analyzed using probabilistic tractography as implemented in the FMRIB Software Library, where FMRIB stands for Functional Magnetic Resonance Imaging of the Brain (www.fmrib.ox.ac.uk/fsl/).
The anatomic network of the cortex and subcortex is measured using Diffusion Tensor Imaging (DTI) and subsequently divided into 90 predefined regions according to the Automated Anatomical Labeling (AAL) atlas \cite{TZO02}. 
Each node of the network corresponds to a brain region. 
Indirect information of the white matter fibers connecting different brain regions is provided by diffusion-weighted Magnetic Resonance Imaging (dMRI) measuring the preferred diffusion direction in each voxel of the brain. Probabilistic tractography then provides for each voxel a set of $n_s = 5\,000$ streamlines, simulating the possible white matter fiber tracts. 
A coefficient $P_{kj}$ giving the connectivity probability from the $k$-th to the $j$-th region is introduced by the proportion of streamlines connecting voxels in region $k$ to voxels of region $j$ on the condition that they originate in region $k$. 
Thus a weighted adjacency matrix of size $90 \times 90$, with node indices $k \in N = \{1,2,...,90\}$ is constructed. Finally, the connectivity between every two regions is averaged over 20 subjects yielding the average empirical structural brain network $\mathbf{A} = \{A_{kj}\}$.
The pipeline for constructing the structural network has been adopted from previous study of differences in connectivity patterns between healthy subjects and schizophrenia patients~\cite{CAB13b}.
Note that in contrast to the original AAL indexing, where sequential indices correspond to homologous brain regions, the indices in Fig.\,\ref{fig.1} are rearranged such that $k \in N_L = \{1, 2, ... ,45\}$ corresponds to left and $k \in N_R = \{46, ... ,90\}$ to the right hemisphere. 
Thereby the hemispheric structure of the brain, i.e., stronger intra-hemispheric coupling compared to inter-hemispheric coupling, is highlighted (Fig.\,\ref{fig.1}a). Note that there is a very slight structural asymmetry of the two brain hemispheres.\\
%Node dynamics:
Each node corresponding to a brain region is modeled by the FitzHugh-Nagumo (FHN) model, a paradigmatic model for neuronal spiking  \cite{FIT61,NAG62}. Note that while the FitzHugh-Nagumo model is a simplified model of a single neuron, it is also often used as a generic model for excitable media on a coarse-grained level. Thus the dynamics of the network reads:
\begin{subequations}
\begin{align}
\epsilon \dot{u}_k = &u_k - \frac{u_k^3}{3} - v_k \nonumber \\
                    &+ \sigma \sum_{j \in N_\text{H}} A_{kj} \left[ B_{uu}(u_j - u_k) + B_{uv}(v_j - v_k) \right]  \\
                    &+ \varsigma \sum_{j \notin N_\text{H}} A_{kj} \left[ B_{uu}(u_j - u_k) + B_{uv}(v_j - v_k) \right], \nonumber \\
\dot{v}_k = &v_k + a \nonumber \\
 & + \sigma \sum_{j \in N_\text{H}} A_{kj} \left[ B_{vu}(u_j - u_k) + B_{vv}(v_j - v_k) \right]  \\
 & + \varsigma \sum_{j \notin N_\text{H} } A_{kj} \left[ B_{vu}(u_j - u_k) + B_{vv}(v_j - v_k) \right], \nonumber
\end{align}
\label{eq.1}
\end{subequations}
with $k \in N_\text{H}$ where $N_\text{H}$ denotes either the set of nodes $k$ belonging to the left ($N_L$) or the right ($N_R$) hemisphere, and $\epsilon = 0.05$ describes the timescale separation between fast activator variable or neuron membrane potential $u$ and the slow inhibitor or recovery variable $v$ \cite{FIT61}. Depending on the threshold parameter $a$, the FHN model may exhibit excitable behavior ($\left| a \right| > 1$) or self-sustained oscillations ($\left| a \right| < 1$). We use the FHN model in the oscillatory regime and thus fix the threshold parameter at $a=0.5$ sufficiently far from the Hopf bifurcation point. The emerging dynamics for an isolated FHN oscillator is displayed in Fig.\,\ref{fig.1}c. The coupling within the hemispheres is given by the intra-hemispheric coupling strength $\sigma$ while the coupling between the hemispheres is given by the inter-hemispheric coupling strength $\varsigma$. The interaction scheme between nodes is characterized by a rotational coupling matrix:
\begin{align}
\mathbf{B} = 
\begin{pmatrix}
B_{uu} & B_{uv} \\
B_{vu} & B_{vv}
\end{pmatrix}
=
\begin{pmatrix}
\text{cos}\phi & \text{sin}\phi \\
-\text{sin}\phi & \text{cos}\phi
\end{pmatrix},
\end{align}
with coupling phase $\phi = \frac{\pi}{2} - 0.1$, causing primarily an activator-inhibitor cross-coupling. This particular scheme was shown to be crucial for the occurrence of chimera states in ring topologies \cite{OME13} as it reduces the stability of the completely synchronized state. 

%-----------------------------METHODS-----------------------------------------------------------------------
\section{Methods}
We explore the dynamical behavior by calculating the mean phase velocity ${\omega_k = 2\pi M_k/\Delta T}$ for each node $k$, where $\Delta T$ denotes the time interval during which $M$ complete rotations were realized. Throughout the paper we use $\Delta T = 5\,000$. Furthermore we introduce hemispheric measures that characterize the degree of synchronization of the sub-networks. First, the hemispheric mean phase velocity is: 
\begin{align}
{\langle \omega \rangle_H = \frac{1}{45}\sum_{k \in N_\text{H}} \omega_k},    
\end{align} 
where $H$ denotes either the left ($H=L$) or right ($H=R$) hemisphere. Thus $\langle \omega \rangle_H$ corresponds to the mean phase velocity averaged over the left or right hemisphere, respectively. To quantify the dynamical difference between the left and right hemisphere we use the difference between these hemispheric mean phase velocities $\Delta \omega = \langle \omega \rangle_R - \langle \omega \rangle_L$.\\ 
Second, the hemispheric Kuramoto order parameter:
\begin{align}
{R_H(t) = \frac{1}{45} \left| \sum_{k \in N_\text{H}} \text{exp}[i \theta_k(t)]\right|},
\end{align}
is calculated by means of an abstract dynamical phase $\theta_k$ that can be obtained from the standard geometric phase ${\tilde{\phi}_k(t) = \text{arctan}(v_k/u_k)}$ by a transformation which yields constant phase velocity $\dot{\theta}_k$. For an uncoupled FHN oscillator the function $t(\tilde{\phi}_k)$ is calculated numerically, assigning a value of time $0<t(\tilde{\phi}_k)<T$ for every value of the geometric phase, where $T$ is the oscillation period. The dynamical phase is then defined as $\theta_k=2 \pi t(\tilde{\phi}_k)/T$, which yields $\dot{\theta}_k = \text{const}$. Thereby identical, uncoupled oscillators have a constant phase relation with respect to the dynamical phase. Fluctuations of the order parameter $R_H$ caused by the FHN model's slow-fast time scales are suppressed and a change in $R_H$ indeed reflects a change in the degree of synchronization. The Kuramoto order parameter may vary between 0 and 1, where $R_H=1$ corresponds to complete phase synchronization, and small values characterize spatially desynchronized states. Finally, we use the spatial correlation coefficient introduced by Kemeth et al.\ \cite{KEM16}:
\begin{align}
\label{corrcoeff}
    g_0(t) = \sqrt{\int_0^{\delta} g(t, D)dD},
\end{align}
that measures the relative amount of synchronized oscillators. It is defined in terms of the normalized probability density function $g(t, D)$, which is calculated as the probability of finding a distance $D$ among all pairwise distances $\{D_{kj}\}= \{\left| e^{i\theta_j} - e^{i\theta_k} \right|\}$ between the states of all oscillators $k,j$, and generalizes the local curvature in systems with a spatial dimension. This distance is calculated using the dynamical phase on the unit cycle where the maximum distance of two oscillators is $D_{\textsl{\textrm{max}}} = 2$. For complete phase synchronization the distance between each pair of oscillators vanishes, i.e., $D = 0$ and $g(t, D) = \delta(D)$, hence $g_0(t)=1$, while a totally incoherent system gives a value of $g(t,0) = 0$, hence $g_0(t)$ is small. Two oscillators are considered spatially correlated if their distance is smaller than some threshold $\delta = 0.01 D_{\textsl{\textrm{max}}}$. The square root in Eq.\,\eqref{corrcoeff} arises because by taking all pairwise distances, the probability of oscillators $i$ and $j$ both being in the synchronous cluster is proportional to the square of the number of synchronous oscillators.
%-----------------------------RESULTS 1-----------------------------------------------------------------------
\section{Dynamical asymmetry}
\begin{figure}
    \onefigure[scale=0.5]{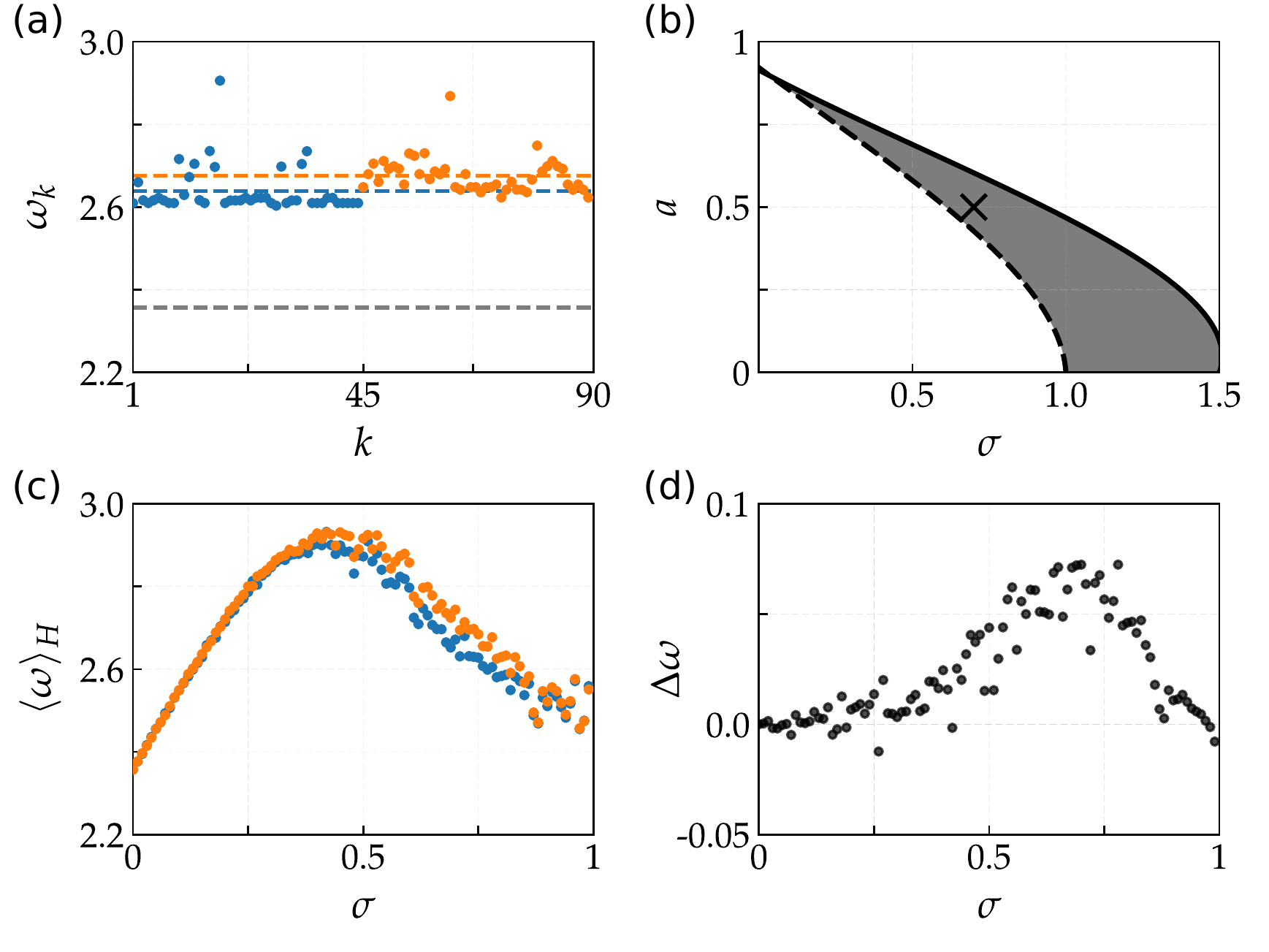}
    \caption{(color online) Asymmetry scenarios in the brain network: (a) Mean phase velocity $\omega_k$ for each node $k$ (dots) and spatially averaged hemispheric mean phase velocity $\langle \omega \rangle_H$ (dashed lines) for coupling strength $\sigma = \varsigma = 0.7$. The color code highlights the left (dark blue) and right (light orange) hemisphere. The gray dashed line at $\omega \approx 2.4$ denotes the mean phase velocity for the uncoupled system. (b) Numerically calculated critical coupling strength in the ($\sigma,a$) plane for the transition between incoherence and frequency synchronization using the average brain network (solid line, $\sigma=\varsigma$) and isolated hemispheres (dashed line, $\varsigma = 0$). The cross denotes the parameters ($\sigma=0.7$, $a=0.5$) used in panel (a). (c) Hemispheric mean phase velocities $\langle \omega \rangle_H$ as a function of the coupling strength $\sigma=\varsigma$, color code as in panel (a). (d) Difference between left and right hemispheric mean phase velocity $\Delta \omega$ as a function of the coupling strength $\sigma=\varsigma$. The difference assumes a maximum at $\sigma \approx 0.7$. Other parameters: $\epsilon = 0.05$, $a=0.5$, $\phi = \frac{\pi}{2} - 0.1$.}
    \label{fig.2}
\end{figure}
\begin{figure}
    \onefigure[scale=0.5]{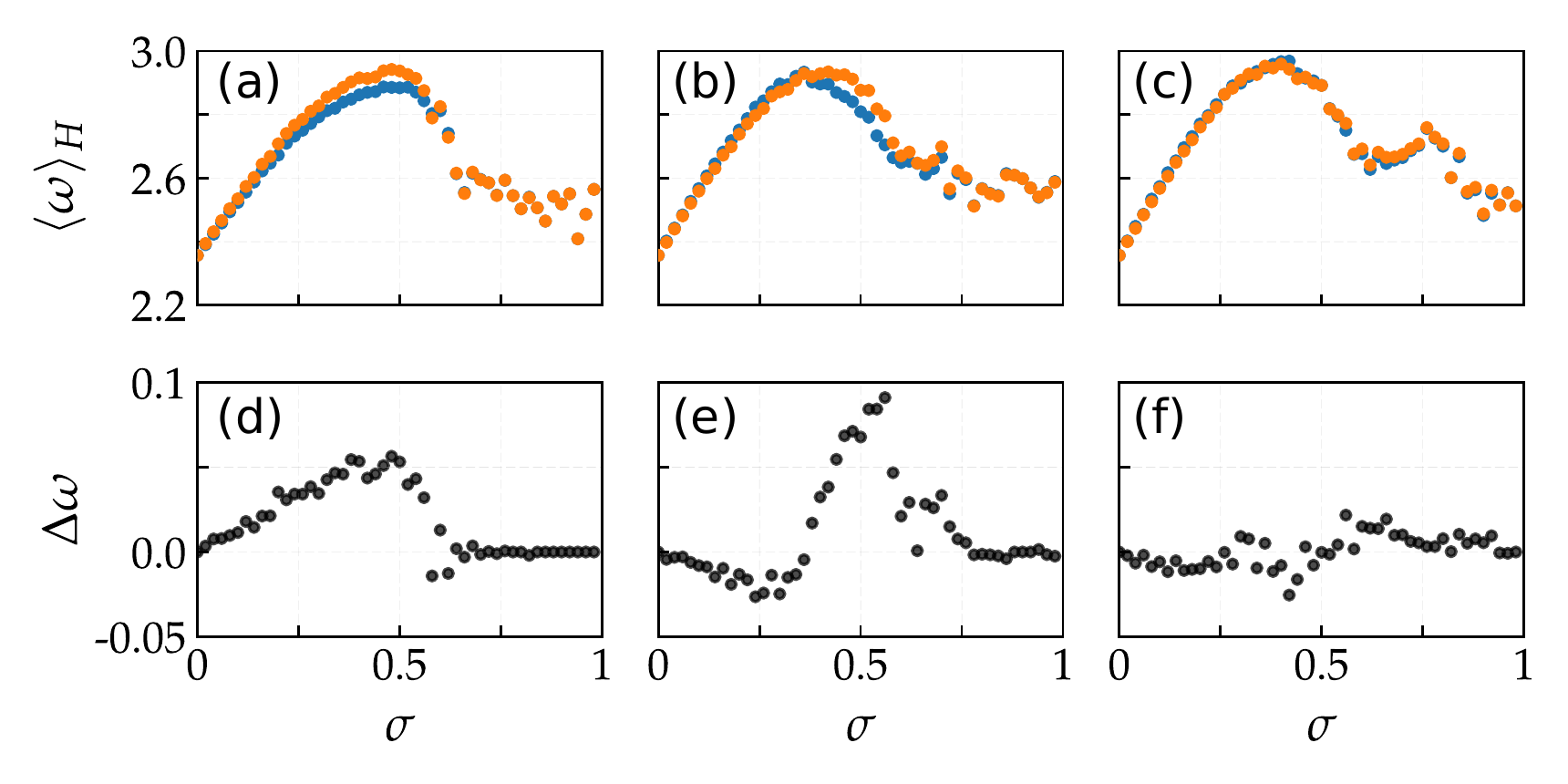}
    \caption{(color online) Classification of the transition between incoherence and frequency synchronization by means of the hemispheric mean phase velocities $\langle \omega \rangle_H$ (a)-(c) and their difference $\Delta \omega$ (d)-(f) as a function of the coupling strength $\sigma=\varsigma$. In 20 individual brain networks three transition scenarios are distinguished, displayed in panels (a, d), (b, e), and (c, f), respectively, each scored 10, 7, and 3 times, respectively. Other parameters as in Fig.\,\ref{fig.2}.}
    \label{fig.3}
\end{figure}

We investigate dynamical asymmetries emerging from the slight structural asymmetry of the brain hemispheres. Figure~\ref{fig.2} shows how the different measures lead to the observation of a dynamical asymmetry with respect to the hemispheres of the average empirical structural brain network. Figure~\ref{fig.2}a displays the node-wise mean phase velocity $\omega_k$ for an intermediate coupling strength $\sigma = \varsigma = 0.7$ with random initial conditions. Note that the oscillators split into two visually well distinguishable communities that coincide with the hemispheres of the brain network and have different hemispheric mean phase velocities $\langle \omega \rangle_H$. The left and right hemispheric mean phase velocities $\langle \omega \rangle_L$ and $\langle \omega \rangle_R$ and their difference $\Delta \omega$ vs $\sigma = \varsigma$ are displayed in Fig.\,\ref{fig.2}c and \ref{fig.2}d, respectively. The values are calculated for one hundred different coupling strengths with $0 < \sigma \leq 1$ and step-size $\Delta \sigma = 0.01$. For every coupling strength an average over ten simulations with different sets of random initial conditions is plotted. For coupling strength $\sigma=\varsigma \ge 1$ the system enters the frequency-synchronized regime, while phase-synchronization measured by the Kuramoto order parameter sets in only later at $\sigma=\varsigma \approx 4.85$. It turns out that the difference $\Delta \omega$ assumes a maximum at $\sigma \approx 0.7$ and subsequently decreases again as both hemispheres enter the frequency-synchronized regime. However, these differences between left and right hemisphere do not imply different dynamical regimes in the sense of a partial synchronization pattern consisting of a desynchronized and a synchronized hemisphere, like in a chimera pattern. Nevertheless it can clearly be concluded that the network dynamics reflects the slight structural asymmetry. Fig.\,\ref{fig.2}b depicts the critical coupling strength for the transition between incoherence and frequency synchronization for a wider range of parameters in the ($\sigma,a$) plane by a solid line for the coupled network with $\sigma=\varsigma$, and by a dashed line for the isolated hemispheres ($\varsigma = 0$). The additional coupling between the hemispheres leads to a higher threshold value $\sigma_c$ for frequency synchronization.\\
So far, we have used an averaged empirical matrix to detect a dynamical asymmetry. For a deeper insight it is important to consider all twenty available empirical structural brain networks individually. In all, we observe one of three transition scenarios from incoherence ($0 < \sigma < 1$) to frequency synchronization ($\sigma > 1$ for $a=0.5$) with increasing coupling strength, as shown in Fig.\,\ref{fig.3}. They are distinguished by the difference of the hemispheric mean phase velocities $\Delta \omega$ exhibiting either a pronounced single maximum (d), or a (negative) minimum followed by a pronounced maximum (e), or essentially no dynamical asymmetry at all (f). However, in most cases ($17$ out of $20$) a dynamical asymmetry was measurable by means of $\Delta \omega$.\\
In the following we analyze to which extent the dynamical asymmetry can be attributed to the structural asymmetry of the network by introducing a structural asymmetry parameter $\rho$ with $0 \le \rho \le 1$ that allows for a continuous tuning between the original structural brain network and a fully symmetrized network, in the sense that both hemispheres are identical. We introduce the coupling matrix elements of a network interpolating between asymmetric and symmetric hemispheres by:
\begin{align}
A^*_{kj} = \rho A_{kj}  + (1-\rho)\overline{A}_{kj}, \quad \rho \in \left[ 0,1 \right]
\end{align}
with $\overline{A}_{kj} = \frac{1}{2} \left(A_{kj} + A_{k+45, j+45}\right)$, where all indices are taken modulo $90$. The resulting matrix $\{A^*_{kj} \}$ describes identical hemispheres if $\rho = 0$ and coincides with the original empirical matrix if $\rho = 1$. 
\begin{figure}[h]
\onefigure[scale=0.5]{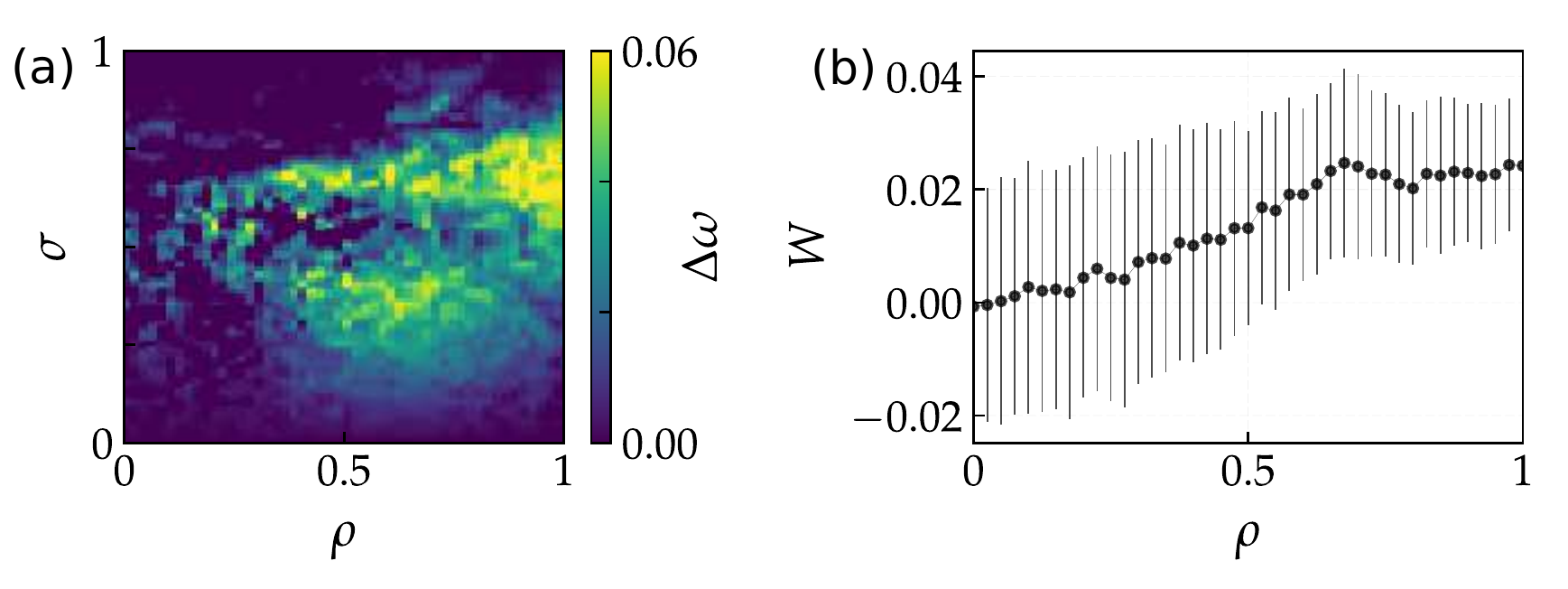}
\caption{(color online) (a) Hemispheric difference $\Delta \omega$ as a function of coupling strength $\sigma=\varsigma$ and structural asymmetry parameter $\rho$. The difference builds up as the structural asymmetry increases. (b) Dynamical asymmetry parameter $W$ as a function of $\rho$. Bars denote standard deviation error of the mean with respect to 20 different realizations of the initial conditions. Other parameters as in Fig.\,\ref{fig.2}. }
\label{fig.4}
\end{figure}
We observe that the dynamical asymmetry, expressed by the hemispheric difference of the mean phase velocities $\Delta \omega $, builds up as the structural asymmetry parameter increases (Fig.\,\ref{fig.4}a). The dynamical asymmetry is most pronounced for an intermediate coupling strength, not to small, but also not too close to the threshold of frequency synchronization. By integrating
\begin{align}
    W = \int_0^1 d\sigma \Delta \omega 
\end{align}
we obtain a dynamical asymmetry parameter $W$ which is indeed almost linearly correlated with the structural asymmetry parameter $\rho$. The Pearson correlation coefficient of these two measures is $r_{\rho, W} = 0.96$. The dynamical asymmetry parameter increases linearly with $\rho$ up to a certain degree of structural asymmetry, but then saturates and does not increase further. This means that even the slightest structural asymmetry results already in a slight dynamical asymmetry, i.e., there is no threshold behavior. However, a slight dynamical asymmetry here does not induce an immediate symmetry breaking as known from critical phenomena. The increase of dynamical asymmetry instead first increases linearly with the structural asymmetry. Beyond this regime of linear response, the dynamical asymmetry does not increase further if the structural asymmetry increases beyond a certain degree, and the real empirical structural asymmetry seems to be just closely above that value corresponding to saturation of sensitivity. These results are consistent with our hypothesis that both unihemispheric and bihemispheric sleep can be possible dynamical states of the same network.
 
%-----------------------------RESULTS 2-----------------------------------------------------------------------
\section{Partial synchronization}
\begin{figure}
\onefigure[scale=0.5]{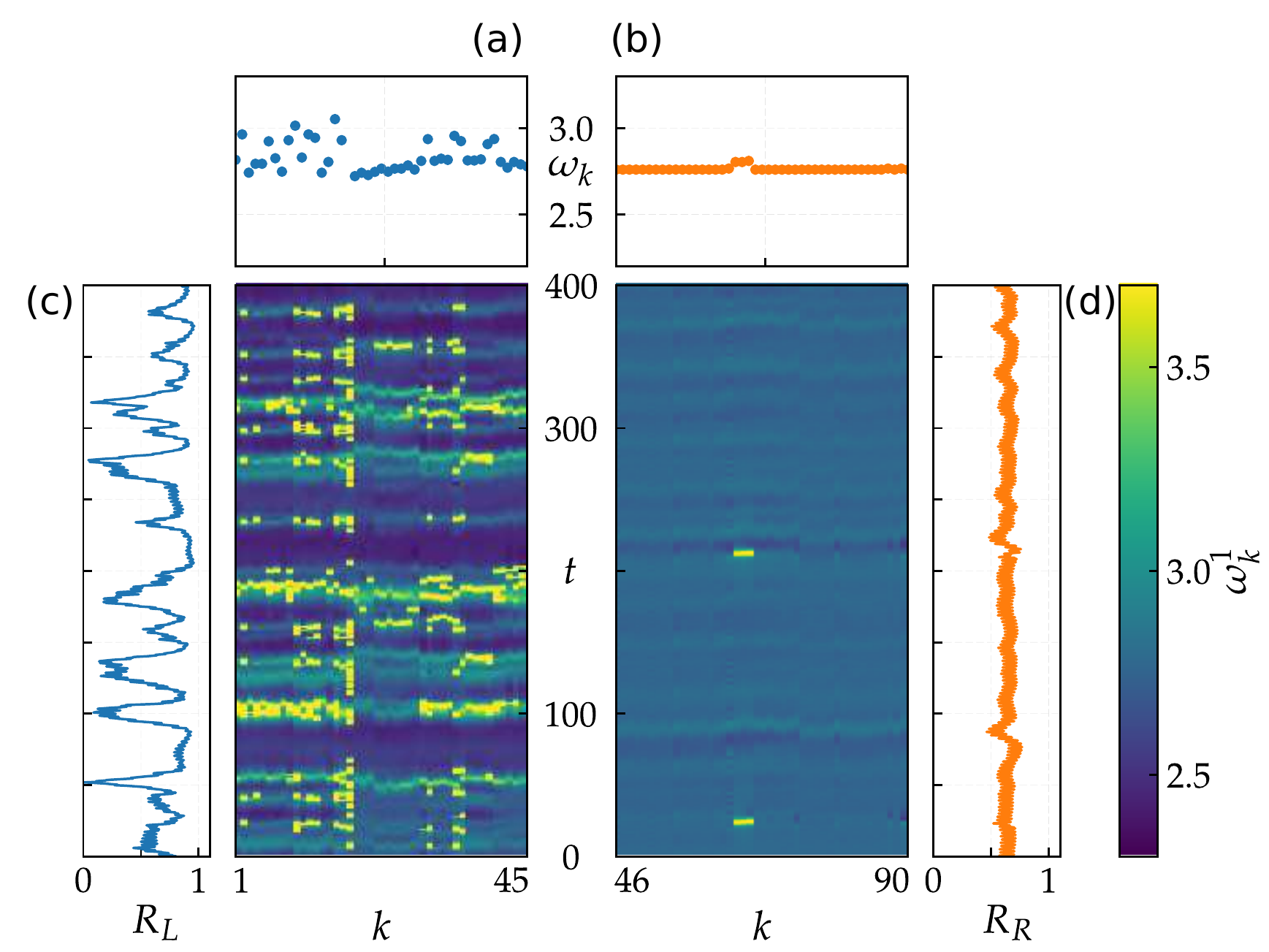}
\caption{(color online) Partial synchronization pattern for $\sigma = 0.70$, $\varsigma = 0.15$ with low and high degree of synchronization in the left (a, c) and right (b, d) hemisphere, respectively. (a),(b) Mean phase velocity profiles $\omega_k$. (c),(d) inner panels: space-time plots of node-wise phase velocity $\omega^1_k$ averaged over a single oscillation, outer panels: hemispheric Kuramoto order parameter $R_{L,R}$ as a function of time $t$. Other parameters as in Fig.\,\ref{fig.2}.}
\label{fig.5}
\end{figure}

\begin{figure}
\onefigure[scale=0.5]{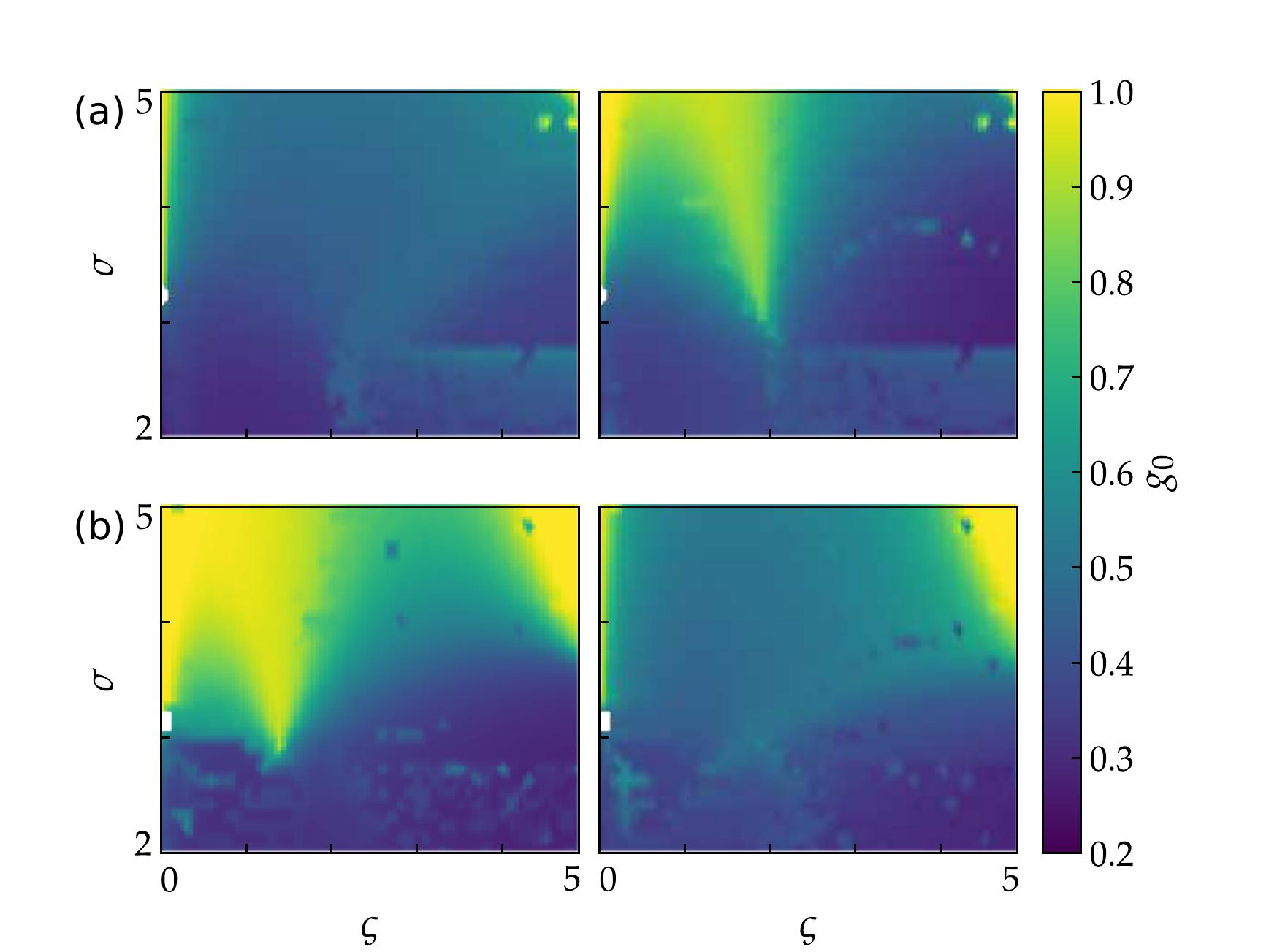}
\caption{(color online) Temporal mean of the spatial correlation coefficient $g_0$ as a function of the intra- and inter-layer coupling strengths $\sigma$, $\varsigma$ in the regime of strong coupling (phase or frequency synchronization). Left and right panels correspond to the left and right hemisphere, respectively. (a) and (b) depict the two possible states of the bistable system. Other parameters as in Fig.\,\ref{fig.2}.
}
\label{fig.6}
\end{figure}

To achieve partial synchronization patterns we consider the inter-hemispheric coupling strength $\varsigma$ as an independent parameter that allows us to reduce the coupling between the hemispheres. This is motivated by the presumption that sleeping with one hemisphere at a time requires a certain degree of hemispheric separation \cite{RAT00}. All other parameters remain unchanged. \\
We analyze the parameter regime where the previously used average empirical structural brain network with identical inter- and intra-hemispheric coupling strength $\varsigma = \sigma$ exhibits qualitatively different behavior from that with separated hemispheres $\varsigma = 0$ , i.e., these two cases correspond to different dynamical regimes. For both cases we have numerically determined the critical intra-hemispheric coupling strength $\sigma_c$ for which the system engages into the frequency synchronized regimes, see Fig.\,\ref{fig.2}b. As $\varsigma = 0$ leaves us with two disconnected sub-networks, these sub-networks are naturally easier to synchronize. Note that these two disconnected sub-networks technically result in two different critical coupling strengths. However, the difference between these critical values is very small and thus negligible. Consider a coupling strength $\sigma$ that lies within the shaded area of Fig.\,\ref{fig.2}b. There, a phase transition with increasing $\varsigma$ must be expected, since the system is frequency synchronized if $\varsigma = 0$, and completely incoherent if $\varsigma = \sigma$. We find that the frequency-synchronized solution indeed breaks down in one hemisphere. This gives rise to the partial synchronization pattern shown in Fig.\,\ref{fig.5} where the left hemisphere is incoherent while the right is frequency-synchronized, except for two small brain regions. This shows up in the space-time plot, in the mean phase velocity profile, and in the hemispheric Kuramoto order parameter (although there is no perfect phase synchronization resulting in $R_R<1$). Note that the incoherent, left hemisphere occasionally exhibits a high degree of synchronization that, in contrast to the right hemisphere, is unstable and vanishes after a short while. In general, partial synchronization patterns where different dynamical regimes occur in the two hemispheres can be found whenever a phase transition with respect to $\varsigma$ is expected, i.e., in the shaded region of Fig.\,\ref{fig.2}b. \\
Finally, we analyze the transition from frequency to phase synchronization which occurs at much higher coupling strengths than shown in Fig.\,\ref{fig.2}b, e.g., for $a=0.5$ at $\sigma=\varsigma=4.8$ and at $\sigma=3.4$ for $\varsigma=0$. Here we use the temporal mean of the spatial correlation coefficient $g_0$ that is suitable to distinguish between phase ($g_0=1$ ), frequency synchronization ($g_0<1$), and complete incoherence ($g_0=0$). In contrast to the Kuramoto order parameter $R$, $g_0$ provides an arbitrary threshold $\delta$ and thereby gives a more pronounced transition from frequency to phase synchronization. Figure~\ref{fig.6} shows that in the ($\sigma$, $\varsigma$) plane of coupling strengths there exist regimes where a high degree of phase synchronization in one hemisphere coincides with a low degree of phase synchronization in the other (i.e., only frequency synchronization). Further, we find that the degree of phase synchronization expressed by $g_0$ may exhibit non-monotonic behavior as a function of $\varsigma$. To a certain amount this is expected as we have seen before that coupling two hemispheres ($\varsigma \ne 0$) decreases the degree of synchronization, we thus expect a maximum of $g_0$ at $\varsigma = 0$. However, in certain regimes of $\sigma$ we find a subsequent maximum at $\varsigma \approx 2$ which implies that increasing the coupling between the hemispheres does not necessarily decrease the degree of phase synchronization. Furthermore, we find that this subsequent maximum is in principle possible in both hemispheres (indicating bistability, cf. panels (a) and (b)), but is realized by only one hemisphere at a time. A high degree of phase synchronization in one hemisphere thus suppresses phase synchronization in the other. This could be an important mechanism for the occurrence of unihemispheric sleep and should be investigated further. It is interesting that the two hemispheres can exchange their roles as one being phase synchronized and the other being only frequency-synchronized. For very strong $\sigma$ and $\varsigma$ complete phase synchronization of both hemispheres is found (top right corners in Fig.\,\ref{fig.6}). 
%-----------------------------CONCLUSION--------------------------------------------------------------------
\section{Conclusion}
We have investigated the dynamical asymmetry arising from the structural difference between the two brain hemispheres. It has been found that during the transition from complete incoherence to frequency synchronization an asymmetry regarding the degree of synchronization builds up, which can be quantified by the different mean phase velocities averaged over each hemisphere. We have shown that this asymmetry can be attributed to the structural asymmetry of the hemispheres, by introducing an asymmetry parameter which can interpolate between the empirical asymmetric brain network and an artificially symmetrized network. Furthermore, we have varied the inter-hemispheric coupling strength, while keeping the intra-hemispheric coupling strength fixed, to increase the degree of inter-hemispheric separation, ranging from isolated to fully coupled hemispheres. This has resulted in the observation of partial synchronization patterns similar to spontaneously synchrony-breaking chimera states. We have demonstrated that these partial synchronization patterns occur for coupling strengths where the isolated hemispheres are frequency-synchronized while the brain network with equal intra- and inter-hemispheric coupling remains completely incoherent. By tuning the coupling between the hemispheres we have shown that at intermediate inter-hemispheric coupling one hemisphere becomes incoherent, giving rise to a chimera-like partial synchronization pattern.\\ 
These results are in accordance with the assumption that unihemispheric sleep requires a certain degree of inter-hemispheric separation. Moreover, it is known that the brain is operating at the edge of different dynamical regimes. By choosing appropriate coupling parameters, we have reported an intriguing dynamical behavior regarding the transition from frequency to phase synchronization. We observe that in this regime the brain exhibits spontaneous symmetry breaking and bistabilty, where each hemisphere may engage into either of two dynamical states, characterized by a relatively high and low degree of synchronization. However, a high degree of synchronization in one of the hemispheres always coincides with a low degree of synchronization in the other. To sum up, the structural asymmetry in the brain allows for partial synchronization dynamics, which may be used to model unihemispheric sleep or explain the mechanism of the first-night effect in human sleep.

%-----------------------------ACKNOWLEDGMENTS---------------------------------------------------------------
\acknowledgments
\noindent This work was supported by the Deutsche Forschungsgemeinschaft (DFG, German Research Foundation) - Projektnummer - 163436311 - SFB 910 and by project Nr. LO1611 with financial support from the MEYS under the NPU I program. LR is grateful to P. Jiruska and J. Hlinka for their hospitality and discussion during his stay at Prague. We are grateful to Anton\'{i}n \v{S}koch for preparing the example structural connectivity matrices.

%-----------------------------BIBLIOGRAPHY------------------------------------------------------------------
%
%\bibliography{ref}
%\bibliographystyle{prsty-fullauthor}

\end{document}